\documentclass[a4paper, 12pt]{amsart}

\usepackage{amsmath}
\usepackage{amssymb}
\usepackage{amsthm}
\usepackage{graphicx}
\usepackage{url}

\usepackage[margin=1truein]{geometry}

\usepackage{enumitem}

\usepackage{amsthm}
\theoremstyle{plain} 
\newtheorem{theorem}{\indent\sc Theorem}[section] 
\newtheorem{lemma}[theorem]{\indent\sc Lemma}
\newtheorem{corollary}[theorem]{\indent\sc Corollary}
\newtheorem{proposition}[theorem]{\indent\sc Proposition}

\theoremstyle{definition} 
\newtheorem{definition}[theorem]{\indent\sc Definition}
\newtheorem{remark}[theorem]{\indent\sc Remark}
\newtheorem{example}[theorem]{\indent\sc Example}

\usepackage{bm}

\begin{document}

\markboth{Taiki Yamada}
{Instructions for Typing Manuscripts (Paper's Title)}

\title{Network analysis using Forman curvature and Shapley values on 
 hypergraphs}

\author{Taiki Yamada}

\address{Interdisciplinary Faculty of Science and Engineering, Shimane University, Shimane 690-8504, Japan.}
\email{taiki\_yamada@riko.shimane-u.ac.jp}

\pagestyle{plain}


\maketitle

\begin{abstract}
In recent years, network models have become more complex with the development of big data. Therefore, more advanced network analysis is required. In this paper, we introduce a new quantitative measure named combinatorial evaluation, which combines the discrete geometry concept of Forman Ricci curvature and the game theory concept of the Shapley value. We elucidated the characteristics of combinatorial evaluation by proving several properties of this indicator. Furthermore, we demonstrated the usefulness of the concept by calculating and comparing the conventional centrality and combinatorial evaluation for a concrete graph. The code is available at \url{https://github.com/Taiki-Yamada-Math/CombinatorialEvaluation}.
\end{abstract}

\section{Introduction}
\subsection{Background}
Accurately calculating the importance of each vertex in a network is a fundamental problem in network analysis. Centrality has been traditionally used as a quantitative measure, for which various definitions exist. Typical examples include order, eigenvector, mediate, and proximity centrality. Clustering coefficients are similar to centrality. As each has its own advantages and disadvantages, they have been used in different manners depending on the network model. For example, order centrality is used in social networking analysis, eigenvector centrality in web searches, mediation centrality in traffic congestion prediction, proximity centrality in crisis management, and clustering coefficients in neuroscience. However, with the recent development of big data and the increasing variety of data, it has become necessary to analyze more complex networks, and more advanced quantitative indicators are required. The concept of discrete geometry, also known as discrete curvature, has been attracting increasing attention. Discrete curvature is a concept that was introduced around 2010 and plays an important role in analyzing network structures. Various definitions of discrete curvature exist, the most common of which are the Ollivier Ricci curvature that is defined using probability theory and the Forman Ricci curvature that is defined using combinatorics. The Ollivier Ricci curvature is computed from global network information; therefore, it can identify important vertices more accurately than centrality, but it cannot be defined for vertex-weighted networks and is computationally slower. By contrast, the Forman Ricci curvature can be defined for networks with vertex and edge weights and is computationally fast. Recently, the Forman Ricci curvature has also attracted attention as a method for analyzing multilayer networks. It has been applied to risk quantification, which is a central issue in financial economics, and various methods have been proposed. Risk quantification has traditionally been based on the centrality of financial networks, where the vertices are companies and the edges are correlations \cite{barabasi2013network}. However, in recent years, the Ricci curvature on graphs has been attracting attention as a new characterization of markets in financial networks. Although several definitions of Ricci curvature on graphs exist, Samal et al. showed that the Forman curvature best captures the system-level characteristics of the market \cite{samal2021network}.
However, as both discrete curvatures are defined on the edges of the network, they are effective for computing the importance of each edge but not for computing the importance of each vertex. Although the scalar curvature, which is defined by determining the average of the discrete curvatures, can be used to calculate the importance of a vertex, it is not an effective measure because it does not correlate with conventional centrality. Therefore, this paper introduces a new indicator by combining the concepts of discrete curvature and game theory.

Game theory can be classified into two types: cooperative and noncooperative. In cooperative game theory, all players cooperate with one another to increase their utility, thereby forming a coalition and generating worth. Cooperative game theory has been applied to risk management in financial economics \cite{Pressacco2000} and market design in the form of auction \cite{WILSON1992227} and matching theories. It also has potential applications in artificial intelligence and machine learning. The primary focus in the application of these theories concerns sharing the worth distribution that is gained from a cooperation. To this end, we usually consider a transferable utility or TU game. In 1953, Shapley introduced the Shapley value \cite{shapley:book1952} based on the results of one of the most famous investigations regarding this topic. Although it corresponds to a method of fairly distributing the worth gained from a cooperation, it is assumed that no restrictions are placed on the cooperation possibilities of the players. To solve this problem, in 1977, Myerson introduced a graph-restricted game by modifying the original game and defined the corresponding Shapley value, namely the Myerson value \cite{Myerson}. Subsequently, several studies have been conducted concerning the Myerson value. The type of cooperative network is an interesting topic that has hitherto been seldom investigated in the literature. Initially, Myerson considered an undirected graph as a cooperative network. Therefore, Li–Shan considered the Myerson value for a directed graph \cite{LI2020142}, and Nouweland–Borm–Tijs considered the Myerson value for a hypergraph \cite{Vanden}. Another topic of research concerns the characterization of the Myerson value. In 2001, Algaba et al. characterized the Myerson value for union stable structures \cite{Algaba}. In 2004, Gomez et al. calculated the Myerson value of splitting graphs for pure overhead games \cite{Gomez}. In 2012, Kim–Hee introduced various concepts of betweenness and centrality. Moreover, they deduced the relationship between these concepts and the Myerson value \cite{Kim}. In 2016, Beal–Casajus–Huettner studied the efficient egalitarian Myerson value \cite{Beal}. In 2021, Wang–Shan proved the decomposition property for the weighted Myerson value \cite{Wang}. 
Therefore, we can confirm that the allocation rule is an effective measure of the importance of vertices in a network. However, the Myerson value cannot reflect the weight of edges or vertices because it is computed by considering the connections of the graph. In this paper, we introduce a new allocation rule named {\em combinatorial evaluation}.

\subsection{Outline}
The remainder of this paper is organized as follows. Section 2 presents the basic notations of hypergraphs and properties of the Shapley and Myerson values. Section 3 presents the definition and properties of the combinatorial evaluation. Section 4 describes the calculation of the combinatorial evaluation on a specific hypergraph and the comparison of centrality for random graphs. Finally, Section 5 lists the main conclusions drawn from this study and identifies the scope for future research. 

\section{Preliminaries}
\subsection{Notations for graph theory}
A graph is typically represented as a pair, $G=(N,E)$, where $N$ is a set of vertices and $\bar{E}$ is a set of edges connecting pairs of vertices.
In this basic representation, all edges are considered equivalent. However, in many cases, edges have distinct values or significance. For example, in a transportation network, an edge may represent an actual distance or transportation cost. To capture this, an edge-weight function $w: \bar{E} \to \mathbb{R}$ is introduced. A graph that is equipped with such a function, denoted as $G=(N,\bar{E},w)$, is called an \textit{edge-weighted graph}. 
Alternatively, assigning values to vertices rather than edges can provide a more accurate representation in certain contexts. For example, in a transaction network, a vertex weight could represent the market capitalization or size of a company. Similarly, a vertex-weight function $m:N \to \mathbb{R}$ can be defined. A graph with this function, which is $G=(N,\bar{E},m)$, is called a \textit{vertex-weighted graph}. 
A graph incorporating both types of weights, denoted as $G=(N,\bar{E},w,m)$, is termed a \textit{doubly-weighted graph}.

This study uses hypergraphs, which are generalizations of graphs as follows.
\begin{definition}
    Let $N$ be a set of vertices.
    \begin{enumerate}
        \item A \textit{hyperedge} $e$ is a subset of $N$. In particular, if the cardinality of $e$ is 2, then $e$ can be identified with an edge of the graph.
        \item A \textit{hypergraph} is represented as a pair, $H = (N, E)$, where $N$ is a set of vertices and $E$ is a set of hyperedges.
        \item For a vertex $i \in N$, the \textit{neighborhood} of $i$ is defined by
        \begin{eqnarray*}
            E_i := \left\{ e \in E \mid i \in E \right\}.
        \end{eqnarray*}
    \end{enumerate}
\end{definition}

Similar to graphs, weight functions can be defined for hypergraphs. Specifically, $w : E \to \mathbb{R}$ is an edge-weight function and $m : V \to \mathbb{R}$ is a vertex-weight function. A hypergraph, that is denoted by $H=(V,E,w,m)$, is called a \textit{doubly-weighted hypergraph}.

Many concepts that are defined on graphs can also be defined on hypergraphs, with slight caution because the endpoints of a hyperedge are sets, not vertices. A path, which is a fundamental concept in graph theory, is prepared in the hypergraph setting.
\begin{definition}
For any two players $s$ and $t$, a \textit{path} between $s$ and $t$ represents a sequence of players and hyperedges $(s = v_0, e_0, v_1, e_1, \cdots , e_{l-1}, v_l = t)$, where
\begin{eqnarray*}
s \in e_0,\ t \in e_l,\ \mathrm{and}\ v_i \in e_{i-1} \cap e_i,
\end{eqnarray*}
for any $i \in \left\{1, \cdots, l-1 \right\}$. 
\end{definition}

\begin{remark}
A path between $s$ and $t$ means that vertices $s$ and $t$ are connected by hyperedges. If there exists a path between $s$ and $t$, then $s$ and $t$ is called \textit{connected}. Moreover, if any pair of players in $N$ is connected, then $H=(N, E)$ is referred to as \textit{connected}. 
\end{remark}

An induced subgraph, which is another fundamental concept in graph theory, is prepared in the hypergraph setting.
\begin{definition}
Let $H = (N, E)$ be a hypergraph. For a subset $S \subseteq N$, the \textit{induced subgraph} $H[S]$  represents a pair of $(S, E_S)$ that satisfy $E_S = \left\{ e \mid e \in E,\ e \subseteq S \right\}$.
\end{definition}
A maximal connected induced subgraph is called the \textit{connected component} of $H$, and $N/E$ denotes the set of all connected components in $H$. Note that the union of all connected components in $H$ equals $N$.

Please refer to \cite{leal2021forman} for the mathematical significance and detailed properties of the Forman curvature. In this section, we only define the Forman curvature and prove its relationship with the combinatorial evaluation.
\begin{definition}[\cite{leal2021forman}]
Let $H = (N, E, w, m)$ be a doubly-weighted hypergraph. For a hyperedge $e \in E$, the \textit{Forman curvature} is defined by
\begin{eqnarray*}
F(e) = \sum_{k \in e} m(k)\left(1 - \sum_{e_l \in E_k \setminus e}\sqrt{\cfrac{w(e)}{w(e_l)}} \right). 
\end{eqnarray*}
\end{definition}
As can be observed from the definition, the greater the number of edges connecting the edge $e$, the more negative the Forman curvature is. The edges with the most negative values play an important role in graph analysis, but the importance of the vertices on those edges cannot be measured. 

The Forman scalar curvature is one method for calculating the importance of a vertex.
\begin{definition}
Let $H = (N, E, w, m)$ be a doubly-weighted hypergraph. The \textit{Forman scalar curvature} for a vertex $i \in V$ is defined by
\begin{eqnarray*}
FS(i) = \cfrac{\sum_{e \in E_i} F(e)}{\sum_{e \in E_i}w(e)}. 
\end{eqnarray*}
\end{definition}
Intuitively, the importance of a vertex $i$ should depend most on the weight of $i$, but the Forman scalar curvature depends strongly on the weights of the surrounding vertices. Therefore, using the Forman scalar curvature to compute the importance of a vertex is not optimal \cite{Akamatsu}.

\subsection{Notation for game theory}
The TU game represents a pair $(N, v)$, where $N$ denotes a set of vertices and $v: 2^N \to \mathbb{R}$ is a characteristic function with $v(\emptyset) = 0$. 
The set of all TU games is denoted by $TU(N)$. 
Each $S \in 2^N$ is referred to as a \textit{coalition}, and $v(S)$ represents the \textit{worth} of $S$.  
\begin{definition}
Given $(N, v) \in TU(N)$, 
\begin{enumerate}
\item $v$ is called \textit{super additive} if for any $S, T \in 2^N$ with $S \cap T = \emptyset$, $v(S \cup T) \geq v(S) + v(T)$.
\item $v$ is called \textit{sub additive} if for any $S, T \in 2^N$ with $S \cap T = \emptyset$, $v(S \cup T) \leq v(S) + v(T)$.
\end{enumerate}
\end{definition}

In cooperative game theory, the question of how to distribute the gains obtained by all players fairly arises. The Shapley value is a method that quantifies the contribution of each player and distributes it accordingly, that is, it expresses the importance of each player.
\begin{eqnarray*}
    Sh_i (N, v) = \sum_{S \subseteq N \setminus \left\{i \right\}} \cfrac{s! (n - s - 1)!}{n!} (v(S \cup \left\{i \right\}) - v(S)).
\end{eqnarray*}
While many properties are related to the Shapley value, the focus of this study is on the following characterization theorem.
\begin{theorem}[\cite{shapley:book1952}]\label{theorem:Shapley}
    The Shapley value can be uniquely determined as efficient, additive, symmetric, and conforming to the null-player property subject to the following conditions.
    \begin{enumerate}
        \item An allocation rule $f$ on $TU(N)$ is \textit{efficient} if for any $(N, v) \in TU(N)$,
	   \begin{eqnarray*}
	   \sum_{i \in N} f_i(N, v)= v(N).
	   \end{eqnarray*}
        \item An allocation rule $f$ on $TU(N)$ is \textit{additive} if for all $(N, v), (N, w) \in TU(N)$,
	   \begin{eqnarray*}
	   f_i (N, v + w) = f_i (N, v) + f_i (N, w)
	   \end{eqnarray*}
	   for any player $i \in N$.
        \item An allocation rule $f$ on $TU(N)$ is \textit{symmetric} if for any $(N, v) \in TU(N)$,
	   \begin{eqnarray*}
	   f_i(N, v) = f_j (N, v)
	   \end{eqnarray*}
	   for any two players $i, j \in N$ with $v(S \cup \left\{ i \right\}) = v(S \cup \left\{j \right\})$ for all $S \subseteq N \setminus \left\{i, j \right\}$.
        \item An allocation rule $f$ on $TU(N)$ satisfies the \textit{null-player property} if for any $(N, v) \in TU(N)$,
	   \begin{eqnarray*}
	   f_i (N, v) = 0
	   \end{eqnarray*}
	   for any $i \in N$ with $v(S) = v(S\cup \left\{i \right\})$ for all $S \subseteq N$.
    \end{enumerate} 
\end{theorem}
As can be observed from the theorem, the Shapley value is an excellent allocation rule that satisfies the property, but it does not consider the graph structure. One solution to this problem is an allocation rule known as the Myerson value. 
\begin{definition}[\cite{Vanden}]
    Let $H=(N, E)$ be a hypergraph, and $(N, v)$ be a TU game. Then the \textit{Myerson value} $\mu (N, v, E)$ of $H$ can be expressed as
	\begin{eqnarray*}
	\mu_i (N, v, E) = Sh_i (N, v^E) = \sum_{S \subseteq N \setminus \left\{i \right\}} \cfrac{s! (n - s - 1)!}{n!} (v^E(S \cup \left\{i \right\}) - v^E(S)),
	\end{eqnarray*}
    where $v^E (S) = \sum_{T \in S/E_S} v(T)$ for any coalition $S$. 
\end{definition}

As the Myerson value is a type of the Shapley value, it satisfies all properties of Theorem \ref{theorem:Shapley}. In addition, the Myerson value is characterized as follows:
\begin{theorem}[\cite{Myerson}]\label{theorem:Myerson}
    Let $H=(N, E)$ be a hypergraph, and $(N, v)$ be a TU game. The Myerson value represents a unique allocation rule that satisfies the conditions of component efficiency and fairness, where
    \begin{enumerate}
        \item an allocation rule $f$ is \textit{component efficient} if we have
	   \begin{eqnarray*}
	   \sum_{i \in T} f_i (N, v, E) = v(T)
	   \end{eqnarray*}
	   for any connected component $T \in N/E$, and
        \item an allocation rule $f$ is \textit{fair} if we have
	   \begin{eqnarray*}
	   f_i (N, v, E) - f_i (N, v, E \setminus e) = f_j (N, v, E) - f_j (N, v, E \setminus e)
	   \end{eqnarray*}
	   for any $e \in E$ with $i, j \in e$.
    \end{enumerate} 
\end{theorem}
The Myerson value is known as an allocation rule that reflects graph structure, but it only considers the connected components of the graph. It does not depend on the weights of the vertices or edges of the graph, making it difficult to apply to practical networks.

\section{New quantitative indicators}
\subsection{Definition}
We consider a supply chain in the manufacturing industry that would be a connected graph, i.e., we consider a hypergraph in which the vertex set is a company that manufactures parts and the edge set is a company that manufactures the final product. In this case, it is natural to assume that the value of a component differs for each company. Therefore, we define the following characteristic function:
\begin{definition}
\begin{enumerate}
    \item Let $(N, E)$ be a hypergraph. For a hyperedge $e \in E$, a map $v_e: 2^{\# e} \to \mathbb{R}$ is an \textit{edge-characteristic function} with $v_e(\emptyset) = 0$. We define $\bm{v}_E$ as the vector of each edge-characteristic function $\left\{v_e\right\}_{e \in E}$ in a row, and denote the set of all $\bm{v}_E$ as $V^E \subseteq \mathbb{R}^{\# e}$.
    \item A \textit{weighted hypergraph game} is a quintuplet $(N,E,w,m,\bm{v}_E)$, where $H=(N,E,w,m)$ is a doubly-weighted hypergraph and $\left\{(N, v_e)\right\}_{e\in E}$ is a set of TU games. 
\end{enumerate}

\end{definition}

We define a new allocation rule by using edge-characteristic functions.
\begin{definition}
Let $(N,E,w,m,\bm{v}_E)$ be a weighted hypergraph game. For any vertex $i \in V$, the \textit{combinatorial evaluation} $CE_i(\bm{v}_E)$ is defined by
\begin{eqnarray*}
CE_i(\bm{v}_E) &=& \sum_{e \in E_i} Sh_i (e, v_e)\\
&=& \sum_{e \in E_i} \sum_{S \subseteq e \setminus \left\{i \right\}} \cfrac{s! (\# e - s - 1)!}{\# e!}(v_e(S \cup \left\{i \right\}) - v_e(S)).
\end{eqnarray*}
Using the concept of the Shapley value, we define $Sh_i (e, w_e) = 0$ if $i \notin e$.
\end{definition}
Although the vertex and edge weight functions do not appear in the definition of combinatorial evaluation, it is possible to provide an edge-characteristic function $\bm{v}_E$ that considers the weights of the vertices and edges.

\subsection{Characteristic}
In this section, we prove the properties of combinatorial evaluation.
The following proposition is easy to prove from the definition.
\subsubsection{General setting}
\begin{proposition}\label{main1}
    Let $(N,E,w,m)$ be a doubly-weighted hypergraph.The combinatorial evaluation is satisfies the following properties. For any edge-characteristic functions $\bm{v}_E$ and $\bm{v}'_E$,
    \begin{enumerate}
        \item (Efficiency) $\sum_{i \in N} CE_i(\bm{v}_E) = \sum_{e \in E} v_e (e)$.
        \item (Additivity) $CE_i(\bm{v}_E + \bm{v}'_E) = CE_i(\bm{v}_E) + CE_i(\bm{v}'_E)$.
        \item (Component efficiency) $\sum_{i \in T} CE_i(\bm{v}_E) = \sum_{e \in E_T} v_e (e)$ for any connected component $T \in H/E$.
        \item (Null-player property) $CE_i(\bm{v}_E) = 0$ for any $i \in N$ and any $e \in E$ with $v_e (S) = v_e (S \cup \left\{i \right\})$ for any $S \in 2^e$.
    \end{enumerate}
\end{proposition}
\begin{proof}
    Basically, we use the characterization of the Shapley value. To show the efficiency, we calculate the following:
    \begin{eqnarray*}
        \sum_{i \in N}CE_i(\bm{v}_E) = \sum_{i \in N} \sum_{e \in E_i} Sh_i (e, v_e)= \sum_{e \in E} \sum_{i \in e}Sh_i (e, v_e) = \sum_{e \in E}  v_e(e).
    \end{eqnarray*}

    To show the additivity, we calculate the following for two distinct edge-characteristic functions $\bm{v}_E$ and $\bm{v}'_E$:
    \begin{eqnarray*}
    CE_i(\bm{v}_E + \bm{v}'_E) &=& \sum_{e \in E_i} Sh_i (e, v_e+v'_e)\\
    &=& \sum_{e \in E_i} Sh_i (e, v_e) + \sum_{e \in E_i} Sh_i (e, v'_e) = CE_i(\bm{v}_E) + CE_i(\bm{v}'_E).
    \end{eqnarray*}

    Next we show the component efficiency. For simplicity, let $H/E = \left\{T_1, T_2 \right\}$ be the set of all components in $H$. For any $i \in T_1$, we assume that there exists a vertex $j$ such that $j \in T_2 \cap \bigcup_{e \in E_i} e$. Then, there exists a path between $i$ and $j$ since $j \in \bigcup_{e \in E_i} e$. However, this contradicts $j \in T_2$. Thus we have $E_i \subseteq T_1$ for any $i \in T_1$; therefore, we calculate the following.
    \begin{eqnarray*}
    \sum_{i \in T_1} CE_i(\bm{v}_E) = \sum_{i \in T_1} \sum_{e \in E_i} Sh_i (e, v_e) =  \sum_{e \in E_{T_1}} \sum_{i \in e} Sh_i (e, v_e) = \sum_{e \in E_{T_1}} v_e (e).
    \end{eqnarray*}
    
    The null-player property is clear from the definition of combinatorial evaluation. The proof is completed.
\end{proof}

We can observe from this proposition that combinatorial evaluation satisfies some properties that characterize the Shapley and Myerson values. However, the combinatorial evaluation does not satisfy symmetry and fairness. Therefore, we discuss alternative properties to symmetry and fairness. In fact, symmetry does not hold for any vertices $i$ and $j$ because the hyperedges containing i and j are not necessarily coincident.
\begin{lemma}\label{lem:weakly symmetry}
    Let $(N,E,w,m,\bm{v}_E)$ be a weighted hypergraph game. Suppose that there exist two vertices $i, j \in N$ with $E_i = E_j$ such that for any $e \in E_i (= E_j)$, we have
    \begin{eqnarray*}
        v_e (S \cup \left\{ i \right\})= v_e (S \cup \left\{ j \right\}),
    \end{eqnarray*}
    for any subset $S \subseteq e \setminus \left\{i, j\right\}$. Then we have
    \begin{eqnarray*}
        CE_i(\bm{v}_E) = CE_j(\bm{v}_E).
    \end{eqnarray*}
\end{lemma}

\begin{proof}
    By the assumption, for any $e \in E_i (= E_j)$, we have
    \begin{eqnarray*}
        Sh_i (e, v_e) = Sh_j (e, v_e).
    \end{eqnarray*}
    Thus, we have
    \begin{eqnarray*}
        CE_i(\bm{v}_E) = \sum_{e \in E_{i}} Sh_i (e, v_e)= \sum_{e \in E_{j}}Sh_j (e, v_e) = CE_j(\bm{v}_E).
    \end{eqnarray*}
    Then the proof is completed.
\end{proof}
\begin{remark}
In a manufacturing supply chain, the assumption stated in Lemma \ref{lem:weakly symmetry} implies that two companies supply parts to the same customers and that the value of those two companies to all customers is equal.
\end{remark}

Let $CE_i(\bm{v}_E)$ and $CE^e_i(\bm{v}_E)$ denote the combinatorial evaluations of $(N,E,w,m,\bm{v}_E)$ and $(N,E \setminus \left\{e\right\},w,m,\bm{v}_E)$, respectively.
According to the definition of the combinatorial evaluation, for any edge $e_0 \in E$ and any $i \in e_0$, we have
    \begin{eqnarray}\label{eq:prefair}
    CE_i(\bm{v}_E) - CE^{e_0}_i(\bm{v}_E) = Sh_i (e_0, v_{e_0}).
    \end{eqnarray}
By Combining Equation \eqref{eq:prefair} and Lemma \ref{lem:weakly symmetry}, we obtain the following corollary.
\begin{corollary}\label{fair}
    Let $e$ be a hyperedge. For any two vertices $i, j \in N$ with $v_e (S \cup \left\{i \right\}) = v_e (S \cup \left\{j \right\})$ for all $S \subseteq e \setminus \left\{i, j \right\}$, we have
        \begin{eqnarray*}
        CE_i(\bm{v}_E) - CE^e_i(\bm{v}_E) = CE_j(\bm{v}_E) - CE^e_j(\bm{v}_E).
        \end{eqnarray*}
\end{corollary}
Corollary \ref{fair} implies the fairness of the combinatorial evaluation.

\subsubsection{Special setting}
In this section, we clarify the relationship with the Forman curvature by considering the combinatorial evaluation using the following specific edge-characteristic function.
\begin{eqnarray*}
    \bar{v}_e (S) = \left| \sum_{i \in S \cap e}m(i) \left(1 - \sum_{e_l \in E_i \setminus e} \sqrt{\cfrac{w(e)}{w(e_l)}} \right) \right|
\end{eqnarray*}
for any $S \subseteq 2^e$. 
Then we obtain
\begin{eqnarray*}
    \bar{v}_e (e) = |F(e)|.
\end{eqnarray*}
The reason for defining the edge-characteristic function as the absolute value of the Forman curvature is that the more negative the Forman curvature, the more important the edges are, but the more positive the edge-characteristic function, the larger the combinatorial evaluation and the more important the vertex is.
Moreover, for any $S, T \subseteq 2^e$ with $S \cap T = \emptyset$, we have
\begin{eqnarray}\label{eq:additivity}
    \bar{v}_e (S \cup T) = \bar{v}_e (S) + \bar{v}_e (T).
\end{eqnarray}

The relationship between the Forman curvature and the combinatorial evaluation is as follows.
    \begin{lemma}
    For any edge $e \in E$, we have
    \begin{eqnarray*}
        \sum_{i \in e} (CE_i(\bm{\bar{v}}_E) - CE^e_i(\bm{\bar{v}}_E)) = |F(e)|.
    \end{eqnarray*}
    \end{lemma}
\begin{proof}
As similar arguments can be made, only the first equation is proven. According to Equation \eqref{eq:prefair}, for any edge $e \in E$ and any vertex $i \in e$, we have
\begin{eqnarray*}
    CE_i(\bm{v}_E) - CE^e_i(\bm{v}_E) = Sh_i (e, v_e).
\end{eqnarray*}
Thus, by the efficiency of the Shapley value, we obtain
\begin{eqnarray*}
\sum_{i \in e} (CE_i(\bm{\bar{v}}_E) - CE^e_i(\bm{\bar{v}}_E)) =  \sum_{i \in e} Sh_i (e, \bar{v}_e) = \bar{v}_e (e) = |F(e)|.
\end{eqnarray*}
Then the proof is completed.
\end{proof}

This proposition implies that the larger the value of the Forman curvature, which indicates the importance of a hyperedge, the greater the impact on the combinatorial evaluation when that edge is removed. More specifically, the combinatorial evaluation can be expressed as follows:
\begin{theorem}\label{calc}
    For any vertex $i \in N$, we have
    \begin{eqnarray*}
        CE_i(\bm{\bar{v}}_E) = m(i) \left| \sum_{e \in E_i}\sum_{e_l \in E_i \setminus e}\sqrt{\cfrac{w(e)}{w(e_l)}} - \#E_i \right|,
    \end{eqnarray*}
    where $\# E_i$ is the cardinality of $E_i$.
\end{theorem}
\begin{proof}
For any vertex $i$ and any hyperedge $e \in E_i$, by the definition of $w_e$, we have
\begin{eqnarray*}
    \bar{v}_e (\left\{ i \right\}) &=& m(i) \left| 1 - \sum_{e_l \in E_i \setminus e}\sqrt{\cfrac{w(e)}{w(e_l)}}\right|.
\end{eqnarray*}
Using this and Equation \eqref{eq:additivity}, we calculate the following.
\begin{eqnarray*}
    CE_i(\bm{\bar{v}}_E) &=& \sum_{e \in E_i} \sum_{S \subseteq e \setminus \left\{i \right\}} \cfrac{s! (\# e - s - 1)!}{\# e!} (\bar{v}_e(S \cup \left\{i \right\}) - \bar{v}_e(S))\\
    &=& \sum_{e \in E_i} \sum_{S \subseteq e \setminus \left\{i \right\}} \cfrac{s! (\# e - s - 1)!}{\# e!} \bar{v}_e(\left\{i \right\}) \\
    &=& \sum_{e \in E_i} \bar{v}_e (\left\{i \right\}) = -\sum_{e \in E_i}m(i) \left| 1 - \sum_{e_l \in E_i \setminus e}\sqrt{\cfrac{w(e)}{w(e_l)}}\right|\\
    &=&  m(i) \left| \sum_{e \in E_i}\sum_{e_l \in E_i \setminus e}\sqrt{\cfrac{w(e)}{w(e_l)}}- \#E_i \right|.
\end{eqnarray*}
The proof is completed.
\end{proof}

\section{Example}
In this section, we calculate the combinatorial evaluation for each vertex, and compare the result with the Forman scalar curvature. 
\begin{example}\label{ex1}
    $H=(\left\{a,b,c\right\}, \left\{ e = \left\{a,b,c\right\} \right\})$, $w(e) = 1$ and $m(a)=1, m(b)=2, m(c)=3$.
\end{example}
Then we have $F(e) = 6$ and 
\begin{eqnarray*}
    FS(a) = FS(b) = FS(c) = 6
\end{eqnarray*}
since the total number of edges is 1. In addition, by Theorem \ref{calc}, we obtain
\begin{eqnarray*}
    CE_a = 1,\ CE_b = 2,\ CE_c = 3.
\end{eqnarray*}
Thus, we can state that combinatorial evaluation considers the vertex weights.

\begin{example}\label{ex2}
    $V(H)=\left\{v_1, \cdots, v_n\right\}, E(H)=2^V \setminus \left\{\emptyset, \left\{v_1 \right\}, \cdots, \left\{v_n \right\} \right\}$, $w(e) = 1$ for any edge $e \in E$ and $m(v_i) = i$.
\end{example}
Then, as the neighborhood sets of all vertices are equal, the Forman scalar curvature of any vertex $i$ can be expressed as
\begin{eqnarray*}
    FS(i) = \cfrac{\sum_{e \in E_i}F(e)}{2^{n-1}}.
\end{eqnarray*}

In addition, by Theorem \ref{calc}, we obtain
\begin{eqnarray*}
    CE_{v_{i}} = i (2^{n-1}-1).
\end{eqnarray*}
Thus, compared with the results of Example \ref{ex1}, the value of the combinatorial evaluation increases as the number of edges increases, even if the vertex weights are equal.

\begin{figure}[h]
     \centering
     \includegraphics[scale=0.5]{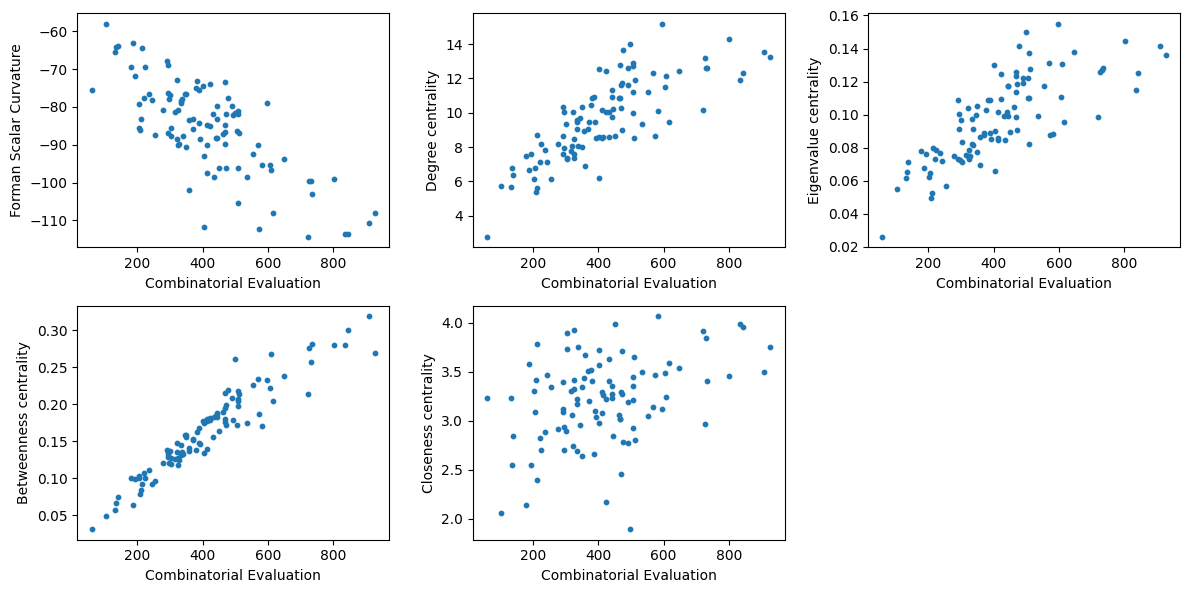} 
     \caption{Scatter plots between combinatorial evaluation and each centrality and between  combinatorial evaluation and the Forman Scalar curvature.}
     \label{data_scat}
\end{figure}

\begin{example}
    Erd\"{o}s--R\'{e}nyi random graphs $G(100,0.2)$. This graph has $100$ vertices, and each possible edge between pairs of vertices appears independently with a probability $0.2$.
\end{example}
As centrality cannot be defined for a vertex-weighted network, only edge-weighted networks are provided. The weight of each edge is assigned randomly in the 0.01 to 1 range. The scatter plots between the combinatorial evaluation and each centrality and between combinatorial evaluation and the Forman Scalar curvature are as follows (see Figure \ref{data_scat}).

\begin{figure}[h]
     \centering
     \includegraphics[scale=0.35]{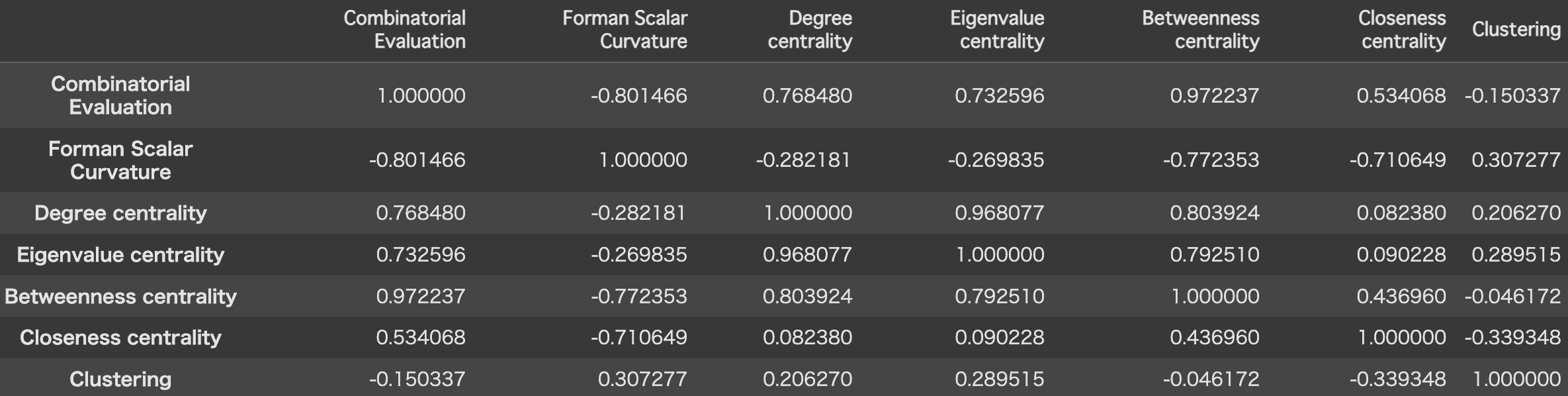} 
     \caption{Correlation between combinatorial evaluation and each centrality (without vertex weights)}
     \label{data_corr}
\end{figure}

In addition, the correlation coefficients are calculated as follows (see Figure \ref{data_corr}).

As these results show, the Forman scalar curvature has almost no correlation with other centralities and is not suitable as a measure of vertex importance. However, our newly introduced combinatorial evaluation has a high correlation with other centralities, and we can conclude that it is effective as a measure of vertex importance. The low correlation between the combinatorial evaluation and clustering coefficients can be attributed to the clustering coefficients not considering the edge weights.

\begin{figure}[h]
     \centering
     \includegraphics[scale=0.35]{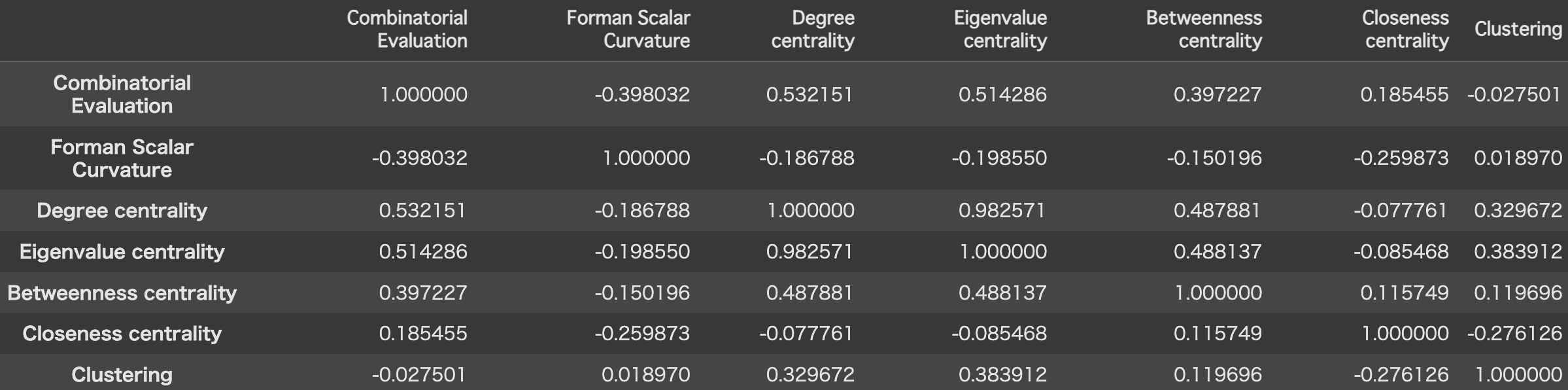} 
     \caption{Correlation between combinatorial evaluation and each centrality (with vertex weights)}
     \label{data_corr1}
\end{figure}

Next, if the weight of each vertex is assigned randomly in the 0.01 to 1 range, the correlation coefficients are calculated as follows (see Figure \ref{data_corr1}). As can be observed from this result, the correlation with the combinatorial evaluation almost disappears because the other centralities do not consider the vertex weights.

\section{Conclusions}
This study aimed to develop an analysis method that is applicable to complex networks. We combined the discrete geometry concept of the Forman Ricci curvature and the game theory concept of the Shapley value to introduce a quantitative measure known as combinatorial evaluation. It not only inherits the characteristics of the Shapley and Myerson values but also appropriately reflects the graph structure and calculates the importance of players. Furthermore, combinatorial evaluation can be defined for networks that have both vertex and edge weights, making it more versatile and applicable to more complex networks than the centrality that is used in traditional network analysis. The implementation of the Python code is expected to be applied in a variety of fields.

In this study, the edge-characteristic function that was used to define combinatorial evaluation was specifically presented and related to Forman Ricci curvature. If a general edge-characteristic function is considered, a more general discussion will be possible.

\section*{{\rm Acknowledgements}}
The author thanks Assoc. Prof. Taisuke Matsubae and Dr. Tomoya Akamatsu for providing important comments. Gratitude is also extended to the anonymous referees for their valuable comments.

\section*{{\rm Funding}}
This work was supported in part by JSPS KAKENHI [grant numbers 21K13800 and 25K17253].

\section*{{\rm Conflict of interest}}
None.

\section*{{\rm Availability of Data and Materials}}
The datasets used and/or analyzed during the current study are available from the corresponding author on reasonable request.

\bibliography{ref}
\bibliographystyle{plain}

\end{document}